\documentclass[letterpaper, 10 pt, conference]{ieeeconf}  
\usepackage[backend=bibtex, style=numeric, sorting=none]{biblatex}
\usepackage{graphicx}
\usepackage{subcaption}
\usepackage{color,soul}
\usepackage{tikz} 
\usepackage{hyperref}
\IEEEoverridecommandlockouts                              
\newcommand\copyrighttext{%
  \footnotesize \textcopyright~2025 IEEE. Personal use of this material is permitted. Permission from IEEE must be obtained for all other uses, in any current or future media,
including reprinting/republishing this material for advertising or promotional purposes, creating new collective works, for resale or redistribution to servers
or lists, or reuse of any copyrighted component of this work in other works.}
\newcommand\copyrightnotice{%
    \begin{tikzpicture}[remember picture,overlay]
    \node[anchor=south,yshift=10pt] at (current page.south) {\fbox{\parbox{\dimexpr\textwidth-\fboxsep-\fboxrule\relax}{\copyrighttext}}};
    \end{tikzpicture}%
}

\title{\LARGE \bf
AUTOFRAME - A Software-driven Integration Framework for Automotive Systems}

\author{
Sven Kirchner$^{1\star}$, Nils Purschke$^{1\star}$, Chengdong Wu$^{1}$, Muhammed Aqib Khan$^{1}$, Divye Dixit$^{2}$, Alois C. Knoll$^{1}$
\thanks{\hspace*{-1em}$^\star$ Equal contribution}%
\thanks{\hspace*{-1em}$^{1}$ Technical University of Munich, Garching, Bayern, Germany}%
\thanks{\hspace*{-1em}$^{2}$ Indian Institute of Technology Mandi, Mandi, Himachal Pradesh, India}%
\thanks{\hrule}
\thanks{\hspace*{-1em}This research was funded by the Federal Ministry of Education and Research of Germany (BMBF) as part of the CeCaS project, FKZ: 16ME0800K.}
}

\begin{document}
\makeatletter
\let\@oldmaketitle\@maketitle
\renewcommand{\@maketitle}{\@oldmaketitle
    \begin{center}
    \includegraphics[width=1.0\textwidth]{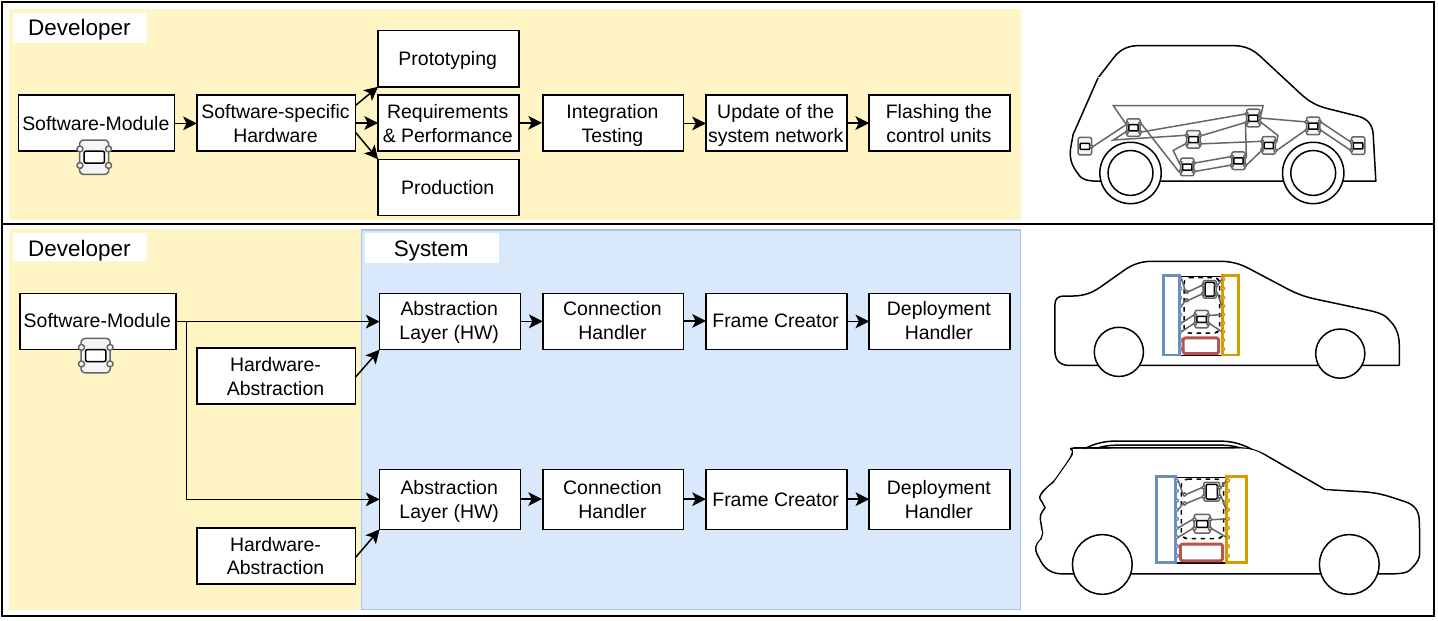}\\
    \captionof{figure}{Comparison of the proposed framework's software integration process (bottom) with the industry standard (top).}
    \label{fig:title}
    \end{center}
}
\makeatother
\maketitle
\copyrightnotice
\thispagestyle{empty}
\pagestyle{empty}

\begin{abstract}
The evolution of automotive technologies towards more integrated and sophisticated systems requires a shift from traditional distributed architectures to centralized vehicle architectures. This work presents a novel framework that addresses the increasing complexity of Software Defined Vehicles (SDV) through a centralized approach that optimizes software and hardware integration. Our approach introduces a scalable, modular, and secure automotive deployment framework that leverages a hardware abstraction layer and dynamic software deployment capabilities to meet the growing demands of the industry. The framework supports centralized computing of vehicle functions, making software development more dynamic and easier to update and upgrade. We demonstrate the capabilities of our framework by implementing it in a simulated environment where it effectively handles several automotive operations such as lane detection, motion planning, and vehicle control. Our results highlight the framework's potential to facilitate the development and maintenance of future vehicles, emphasizing its adaptability to different hardware configurations and its readiness for real-world applications. This work lays the foundation for further exploration of robust, scalable, and secure SDV systems, setting a new standard for future automotive architectures.
\end{abstract}

\addtocounter{figure}{-1} 

\section{INTRODUCTION}
The automotive industry is facing a challenge. Intelligent Transportation Systems (ITS) are becoming increasingly complex due to automated driving functions, vehicle connectivity systems (V2X), and various infotainment solutions. In short, an ever-growing list of functions requires more and more software and computing power in the vehicle \cite{Antinyan2020}. The need for a software and hardware platform that can meet these demands is growing.

However, it has been industry practice for years to add more electronic control units (ECUs) to vehicles as new functions are introduced, resulting in the evolutionary development of an outdated vehicle architecture \cite{Chakraborty2012}. Distributed systems like these require a high amount of coordination, especially when several manufacturers are involved, each working independently on the functionality they provide, only coming together at the end to integrate and merge their results. Each manufacturer uses different hardware, software, and development tools to solve the problems they face. Once a suitable target hardware architecture has been selected, the software heavily depends on it. The application is flashed directly to the ECU, with no means of easily updating or upgrading it, except by manual intervention via a hardware bus. Finally, all the ECUs are hardwired with hundreds of meters of cable \cite{Leen2002}. This situation presents significant maintenance challenges across both hardware and software dimensions. There is no easy way to debug, update, or upgrade the car. Yet these are some of the key challenges facing the car of the future. This will be even more important when the car is fully autonomous and the user can concentrate on other tasks during the journey. In this case, infotainment functions with a much shorter life cycle \cite{6479801} will play a central role in car purchase, as the differentiation between different car models and manufacturers in terms of driving capabilities will disappear.

One proposed solution to the mentioned problems is the central car architecture, where the car's computing power is mainly centralized in a single control unit that runs most or even all of the software \cite{Askaripoor2022}. This simplifies the implementation of the software defined vehicle concept, in which the software defines the functionality of the car rather than the hardware. Centralization makes software development much more dynamic, as updates only need to be applied to one controller rather than all the ECUs involved. In addition, well-coordinated hardware-software co-design is essential to meet the demands placed on the software while providing the developer with a simple and modular development environment \cite{5440056}. An underlying hardware abstraction layer (HAL) decouples the software from the hardware, facilitating agile deployment. The same applies to the integration of new actuator- or sensor-hardware. Instead of having to adapt the whole system software- and hardware-wise, including more wiring and an additional processing ECU, in a centralized architecture, it is only necessary to adapt the sensor and actuator HAL, or if the actuator or sensor is already supported, new hardware can even be added in a plug-and-play fashion.

However, the new architecture also presents challenges. Ensuring safety in such a complex and interconnected system is paramount. As more components rely on the central computing platform, any vulnerability or failure in the system could have widespread implications. But also modularity, scalability, and real-time requirements are fundamental to successfully implementing Software Defined Vehicles using a centralized platform. This work addresses three critical open questions in this research area and is closing the gap of integrating them into a cohesive, operational system: centralized systems' modularity, safety, and scalability across various hardware configurations.

Our \textbf{contributions} (Figure \ref{fig:title}) to research and development for continuous improvement in the area of SDV in centralized vehicle architectures are in detail:
\begin{itemize}
\item Design and implementation of a scalable, modular and safe automotive deployment framework.
\item Design and implementation of a hardware abstraction layer adapting itself automatically according to the vehicle setup.
\item Extending an existing simulation environment to a fully modular and SDV-enabled simulation environment for testing and developing novel software architectures with freely configurable vehicle setups.
\item Application of the shown methodology through an automated driving use case.
\end{itemize}

\section{RELATED WORK}

The concept of centralized vehicle architectures has gained traction within the industry, as highlighted in prior research \cite{7927914}. Efforts to consolidate multiple Electronic Control Units (ECUs) into a single, more advanced ECU \cite{Burkacky2018} and the adoption of open software architecture standards such as AUTOSAR \cite{Fuerst2009} are indicative of a move towards software-defined vehicles with centralized computing architectures. Pioneering this concept, the authors of \cite{Sommer2013} and \cite{Stähle2013} demonstrated the feasibility of safely operating real vehicles via a central computer, addressing the challenges posed by increasing software complexity.

\subsection{Centralized Architecture}
The zonal approach in vehicle design reduces the length and weight of wiring harnesses and offers functional advantages \cite{Maier2023}. Central to the success of centralized E/E architectures is the role of virtualization, which ensures freedom from interference and facilitates straightforward software integration \cite{9337216}. Vehicle systems have evolved from distributed systems to domain-specific and ultimately centralized system architectures. The trend is moving towards replacing microcontrollers with Systems on Chips and scalable central computing platforms \cite{7927914, 9952757}.

\subsection{Software Defined Vehicle}
The concept of defining vehicle architecture through software is well-established in automotive research. Numerous studies, such as \cite{9318389}, have explored adapting software-defined networking principles to automotive applications, often with a focus on developing reconfigurable in-vehicle networks to enhance flexibility and modularity. Initiatives like integrating centralized information and communications technology (ICT) architectures into vehicles \cite{6601320} and building ICT architectures for future electric vehicles \cite{6183198} are also noteworthy.

\subsection{Automotive Middleware}
Various automotive middleware approaches exist, addressing different functional aspects. One of the best known is AUTOSAR \cite{Fuerst2009}. It provides a base run-time environment and base software to execute software units - atomic software blocks - written by the developer. Focusing on extensibility, \cite{yoo2012android} proposed a middleware based on the operating system Android using a master ECU to ensure the installation and modification of automotive applications can be done in a plug-and-play fashion. With security in mind, \cite{bouard2012driving} developed a middleware extension for secure network communication services, enabling the simplified integration of security measurements into overlying applications. Integrating new, unknown devices into an automotive system, is a task requiring an adaptive middleware. The authors of \cite{anthony2007policy} have shown the feasibility of this by adding a discovery and resource management service into their architecture, allowing for an automated integration of additional devices to an existing system after deployment. 

However, an approach that combines all three areas is still lacking. At the same time, we are not aware of any solutions designed specifically for centralized architectures. These have special requirements e.g. for application isolation, because multiple applications run on the same hardware platform. Additionally, the existing solutions mostly rely on static system design and are hard to extend themselves. Therefore, we present a novel framework that addresses these requirements of future intelligent transport systems with centralized architectures.

\section{FRAMEWORK}
\begin{figure}[h]
    \centering
    \includegraphics[width=0.45\textwidth]{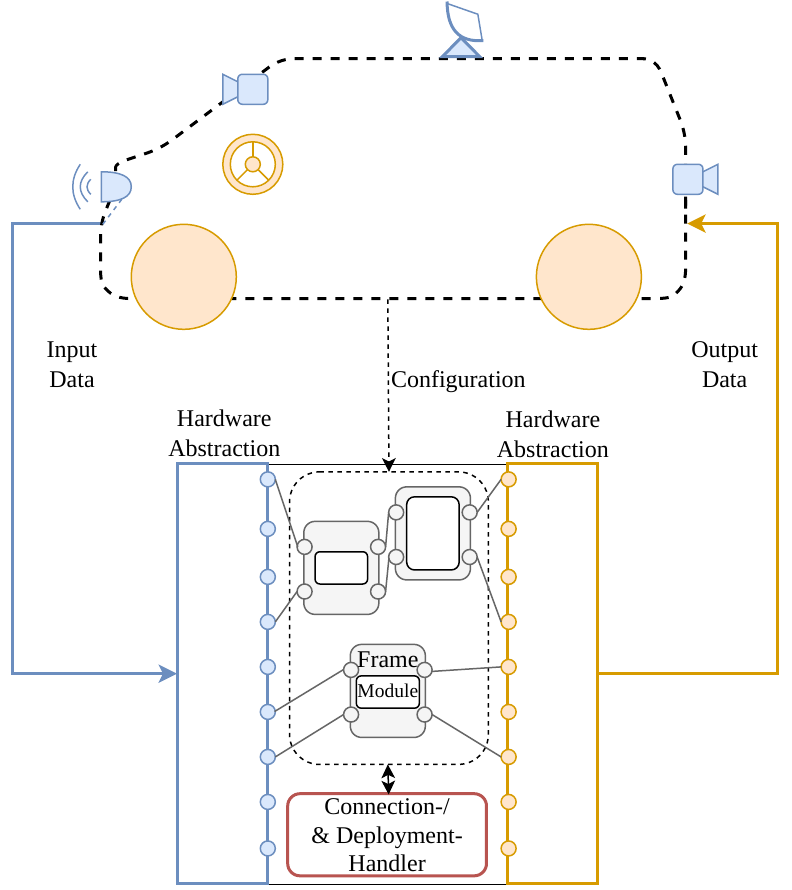}
    \caption{Overview of the integrated system consisting of vehicle, central architecture and framework.}
    \label{fig:overview_integrated_system}
\end{figure}

In response to the increasing diversity of vehicle hardware configurations and functional requirements, we propose a framework for flexible software deployment across different automotive platforms. This framework is fundamentally structured to decouple software from hardware, thereby facilitating its autonomous deployment and integration into the system architecture, which assists in managing escalating complexity. The system consists of the following components, which can be defined, scaled, and modularised on the basis of certain degrees of freedom (Figure \ref{fig:overview_integrated_system}):

\begin{itemize}
\item Vehicle Configuration
\item Application-Modules
\item Hardware-Abstraction
\item Connection-Handler
\item Deployment-Handler
\end{itemize}

\subsection{Vehicle Configuration}
Our framework utilizes a vehicle configuration framework, structured in a standardized JSON file format. This file contains comprehensive information about all vehicle parameters.

Key details in the configuration include the vehicle model and its physical attributes such as weight and tire condition. This data is crucial for vehicle dynamics controllers and similar systems to accurately evaluate the vehicle's performance in various scenarios.

The configuration also includes extensive information about the vehicle's internal devices, including cameras, lidars, radars, engines, and braking systems. Each sensor is detailed by its position and orientation relative to the vehicle, enabling software modules that rely on this data to function effectively. Additionally, sensor-specific details are provided; for example, the image sensor size for cameras or the capture rate for lidars. Parameters like these are essential for algorithms like motion planners, which must take into account sensor placement, size, and other hardware specifics to accurately predict the movement of other road users relative to the vehicle.

Moreover, the configuration includes necessary connection details for each device, such as the device name, manufacturer, connection protocol, network address and port. This information facilitates the establishment of connections to the devices within the vehicle.

\subsection{Application-Modules}
\begin{figure}[h]
    \vspace{1em}
    \centering
    \includegraphics[width=0.4\textwidth]{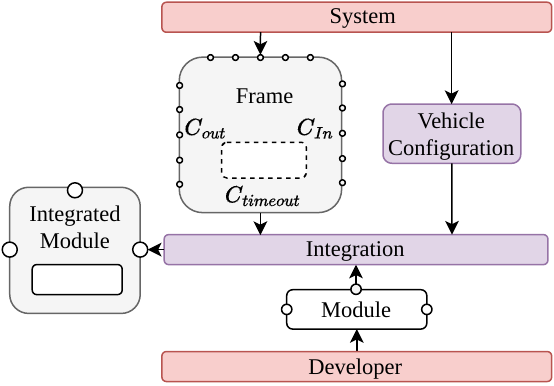}
    \caption{Pre-processing of a developed application for modular integration into the system.}
    \label{fig:module_integration}
\end{figure}

In the provided framework, applications developed by software engineers are the smallest operational units within the system and are referred to as ``modules". The system is designed to offer maximum flexibility in the development of these modules, including an interface to the common programming language Python. These modules are automatically assigned within the system framework (Figure \ref{fig:module_integration}). Each module is incorporated into a frame that establishes standardized interfaces with the overarching system. This frame facilitates essential connections and incorporates interfaces for critical safety functions such as timeout management \(C_{timeout}\) and input/output checks \(C_{in}\), \(C_{out}\). The parametrization of modules is conducted through a configuration file, which is tailored according to the specific vehicle configuration. Information required from other modules, as well as connections necessary for later deployment, can be stored in a designated data path. 

\subsection{Hardware-Abstraction}
Hardware-Abstraction stands as a key challenge in the development of centralized, scalable, and modular vehicle architectures. It allows the execution of software independently from the underlying hardware, sensors, and actuators, facilitating the utilization of the same software across various vehicles and vehicle classes. To achieve hardware independence, a hardware abstraction layer is constructed. Above this layer, standardised interfaces are provided, which are made available to the system and the applications running on it. 

As a starting point, the hardware configuration is provided to identify which sensors can be accessed and which actuators can be controlled by the software that is going to be deployed. Those configuration parameters are used to determine the amount of hardware abstraction modules, their type, configuration, and how to build a connection to the underlying hardware. On the sensor side, those modules produce unified data, namely captures as, e.g., a camera image, which can be received by other modules. On the actuator side, they can be used to control the underlying actuators by sending unified data, namely commands, e.g., a steering angle to them. In this way different sensor and actuator setups in varying vehicle models and categories are addressed.
As data representation follows a standard format, common methods can ensure data consistency across different modules that produce the same outputs. For example, these methods can verify that the steering angle set by a command does not exceed the maximum allowable angle.

\subsection{Connection-Handler}
\begin{figure}[h]
    \vspace{1em}
    \centering
    \includegraphics[width=0.4\textwidth]{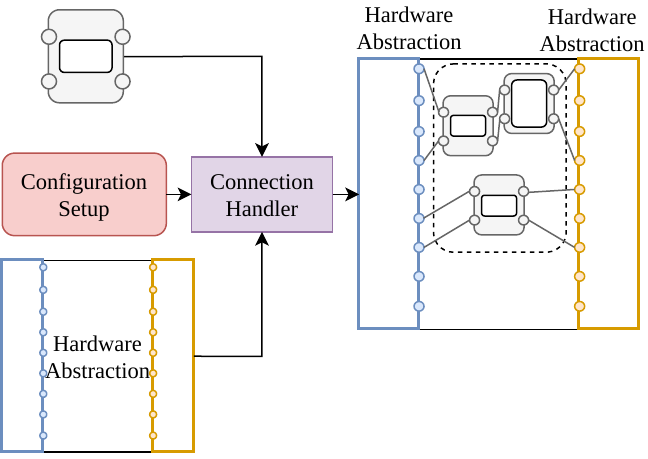}
    \caption{Definition of the communication blueprint in the system via the Connection-Handler.}
    \label{fig:communication_blueprint}
\end{figure}

Standardized communication at both signal and data level allows modules to be strictly separated from each other. In the system architecture, once all the modules have been configured, connections between them are established based on the abstract definitions of their inputs and outputs (Figure \ref{fig:communication_blueprint}). This abstract level of inter-module communication provides a blueprint for the deployment phase. During this phase, concrete connection details such as IP addresses and ports are provided based on this abstract representation of the connections. Therefore a straightforward way of expanding and adapting the communication protocols as needed is supported.

\subsection{Deployment-Handler}
As all modules run on a central compute platform by default, memory and hardware access separation must be ensured so that one module cannot modify the memory of another module or consume all hardware resources. Standardized communication at both signal and data level allows modules to be strictly separated from each other. This is crucial to ensure that if one module fails, all others can continue their work, provided they are not dependent on it. The deployment environment for modules utilizes Docker containers \cite{merkel2014docker} to ensure this isolation. 

Before deployment, module-specific data can be prepared and initialized in a designated directory. This has to be done in a configuration creation method which must be implemented for each module. It provides the path to this directory, along with the vehicle configuration and relevant hardware paths. Subsequently, this method returns a module configuration, encompassing details such as timeout intervals and required external ports. Each module, along with its associated data, is than during deployment automatically copied into a container and executed. 

To isolate applications and enable targeted communication, each module is integrated into a defined frame and deployed within its Docker container. Communication channels are established based on stored information, allowing modules to initialize and commence data exchange. Input data is continuously provided to the modules through the communication interfaces. This is followed by input validation, data processing, output validation and output forwarding, while at the same time timeouts are checked. Standardized data representation ensures consistency and facilitates module exchangeability. Potential errors can be handled by designated methods.

The deployment process can be extended by implementing a provided interface. Through this the framework supports scalability across different hardware platforms and even enables distributed deployment on multiple devices, while ensuring a unified system appearance from a software perspective, thereby facilitating hardware modifications or expansions.

\section{Evaluation}
In the following, the concepts presented in this work will be tested and demonstrated. For this purpose, we have built a scenario generation application using the driving simulation environment CARLA \cite{Dosovitskiy17}, implemented a version of the presented framework in Rust \cite{Klabnik2023}, and written different hardware abstraction components, as well as a lane detector, a motion planner, and a control algorithm in Python that will work as modules.

\subsection{Hardware Setup}
To enhance the efficiency of our experiments, we have implemented a robust computing infrastructure capable of real-time simulation of numerous sensors in parallel (Table \ref{tab:hardware}). Our hardware setup consists of two computers. The first one is the simulation computer. This unit handles all aspects of the simulation, including the driving environment, sensors, actuators, and vehicle physics.

The proposed framework with embedded application modules runs on the second computer and is used to control the vehicle. By separating the simulation and the control unit, a real-world setup is emulated, where the control unit communicates with sensors and actuators, that are separate devices.

Both machines are using Ubuntu 22.04 with a real-time kernel as their operating system.

\begin{table}[h]
\centering
\begin{tabular}{|l|l|l|}
\hline
\textbf{Name} & Simulation Computer & Central Car Computer \\ \hline
\textbf{CPU} & AMD TRP 5995WX & AMD TRP 5955WX \\ 
 & 64 Cores at 2.7 GHz & 16 Cores at 2.7 GHz \\ \hline
\textbf{GPU} & 6 x RTX 4090 24 GB VRAM & RTX 4090 24 GB VRAM \\ \hline
\textbf{RAM} & 256 GB DDR4-3200 RAM & 64 GB DDR4-3200 RAM \\ \hline
\end{tabular}
\caption{Hardware setups - The simulation computer is equipped with six graphics processors to simulate a large number of sensors, while the vehicle computer has much less computing power to emulate a realistic scenario.}
\label{tab:hardware}
\end{table}

\subsection{Simulation}
CARLA allows virtual driving scenarios to be created via the Python Interface (Python API) provided, enabling the direct use of incoming data. However, for the purposes of this work, a solution is needed that allows an easily configurable scenario setup that does not directly process incoming sensor data and control the vehicle, but rather forwards it.

We therefore developed an extension that facilitates the modification of data formats to test different hardware abstraction modules and allows the configuration of sensors, actuators and physics parameters for reuse in different vehicle setups. It also automates the connection to framework modules, while allowing the provision of vehicle setups in an abstract format.

By creating vehicle configurations using specified JSON files, relevant information about the vehicle, like its sensors and actuators, can be loaded at startup and is stored for future use. The application then connects to the CARLA server using the CARLA Python API. Once the connection is established, a freely definable scenario is loaded. For this purpose, a base class is provided which simplifies the definition of such a scenario by allowing the selection of a map and the vehicles to be spawned. The scenario also receives the configuration and can access the loaded data from it. In this way, the vehicles defined in the configuration can be spawned using a method from the scenario base class.

By adding a sensor listener and a vehicle controller, both with predefined base classes, to a vehicle, the sensor data can be received and the car can be controlled. This setup can be easily extended by adding sensor listeners and vehicle controllers, e.g. to provide a specific behaviour of another road user to test the system. As the scenario may need to be dynamically adapted, a method called continuously at each simulation step, one to set up the scenario and another one called on tear down of the scenario are included in the interface.

For the purposes of this experiment, we have implemented a sensor listener and a vehicle controller that sends sensor data via Transmission Control Protocol (TCP) to an IP address and port specified in the configuration files, which is in our case the central car computer. The same applies to the control of the virtual vehicle actuators, although TCP servers are used as endpoints, to which the computer connects.

We used a map consisting of a large single-lane test track, which is specifically designed for the purpose of testing our systems. It includes curves to validate the performance of the detection, planning and control algorithms.

In addition, to further simplify the process and avoid having to provide the framework with the vehicle configuration file for each scenario change, a debug server can be enabled on the simulation extension side, while a corresponding endpoint for the framework is also implemented, which loads and adapts the current vehicle configuration at each scenario change.

\begin{figure*}[htb]
    \vspace{1em}
    \centering
    \includegraphics[width=0.96\textwidth]{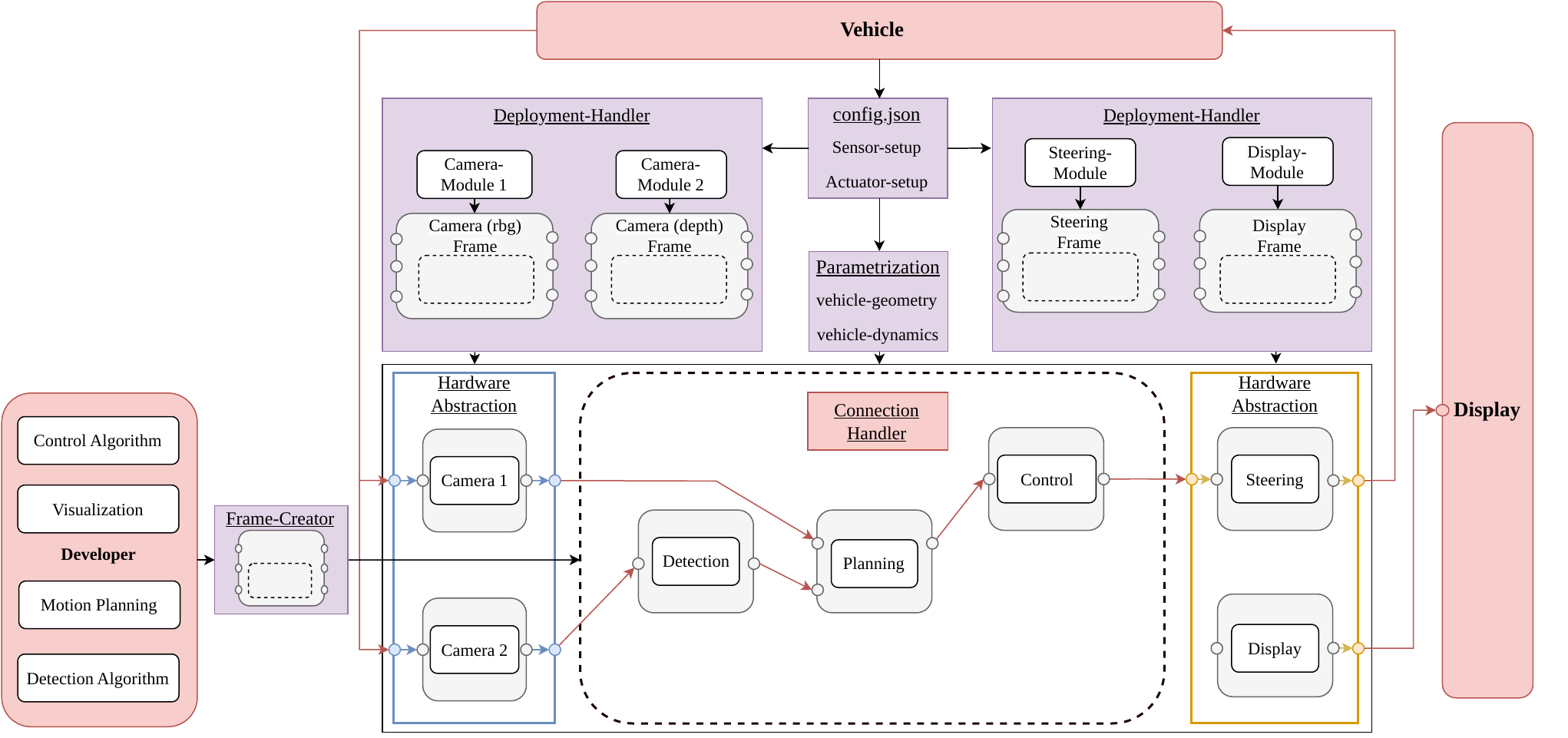}
    \caption{Deployment process for an extended lane following use case (dataflow).}
    \label{fig:deployment_process}
\end{figure*}

\subsection{Modules and Framework}
After the configuration has either been received via the debug server or loaded from the file system, the first step of the framework is to build an appropriate hardware abstraction layer based on the vehicle setup contained in the configuration. This layer abstracts the concrete sensors and actuators from a software perspective. For the evaluation case, a Mercedes C-Class Coupé is used as a vehicle model. The sensor setup consists of an RGB camera and a depth camera, while the actuators are the steering and a display. All components except the display are simulated by CARLA. According to the hardware setup, pre-implemented hardware abstraction modules are automatically used. In accordance with the framework presented, the following applications are now being integrated into the system (Figure \ref{fig:deployment_process}).

Lane Detection:
Only the RGB camera images are needed to detect the lane markings. They are received by the hardware abstraction module of the virtual CARLA rgb camera. Using OpenCV \cite{opencv_library} for edge detection provides the lane markings to the right and left of the lane. The center of the lane is determined by calculating the centers of the right and left lanes at several positions and then smoothing them.

Motion Planning:
The Motion Planner receives the lane center points from the lane detection module as input. To calculate three-dimensional waypoints, depth information is essential. This is obtained from a depth image provided by the depth camera abstraction module. In addition, the cameras optical model is required to convert these points into world coordinates. In detail the camera's field of view, which is specified in the vehicle configuration, within the camera specifications. By converting the two-dimensional track-centered points into three-dimensional world points, the track to be followed is planned.

Control:
We use a combination of proportional-integral-derivative (PID) and model predictive control (MPC) to steer the vehicle along pre-defined waypoints. Using a single-track model, the MPC is the feed-forward part of the controller, while the PID is implemented as the feedback part to compensate for the tracking error results from the simplified vehicle model. The combined controller receives vehicle states, physical parameters and middle points as inputs and calculates the steering angle.

Once the Connection-Handler has created the blueprint for communication, the applications are integrated according to the frames provided in the Deployment Handler.

\subsection{Dynamic Deployment}
\begin{figure*}[ht]
    \vspace{1em}
    \centering
    \includegraphics[width=1\textwidth]{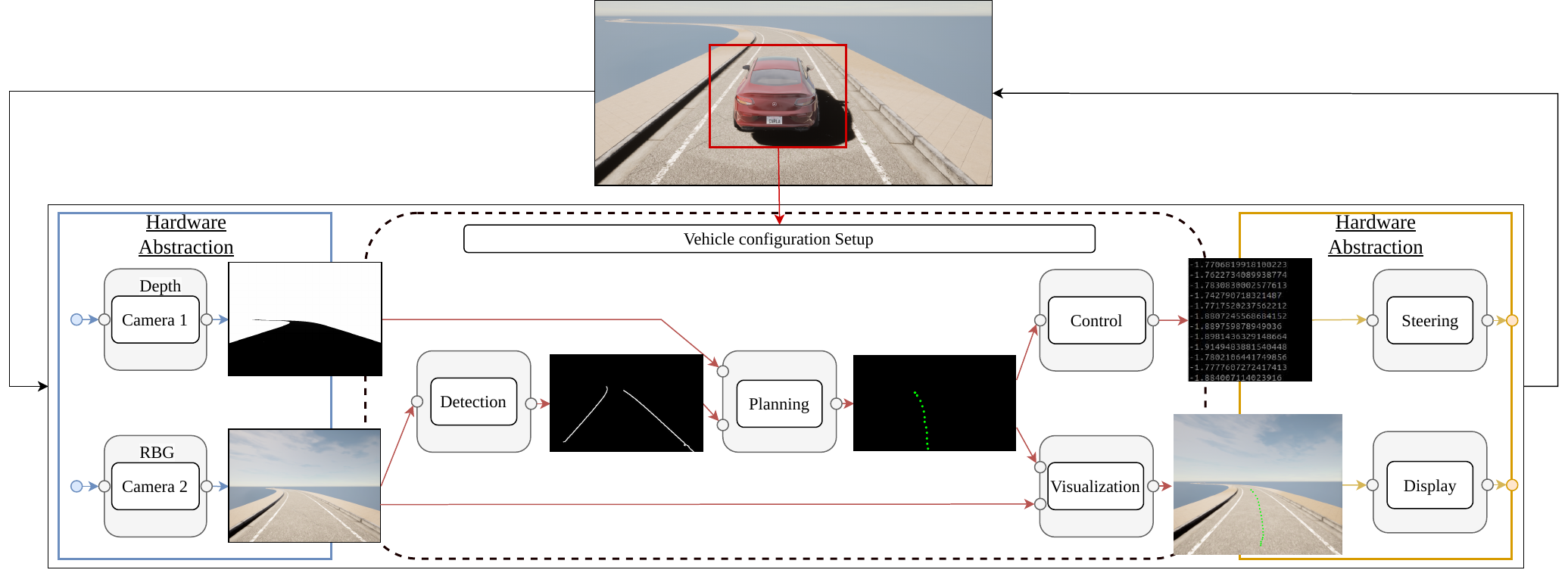}
    \caption{Data flow in the integrated lane following system after completion of the integration process.}
    \label{fig:fully_deployed}
\end{figure*}

To showcase the ease of adding a new module to the system, we developed an additional module which was not yet deployed, a driving trajectory visualization application. It takes the waypoints as well as the rgb camera as an input and outputs the planned vehicle tajectory on a display. To integrate this application into the system, it is simply added to the available vehicle modules. The rest of the process is fully automated. A new connection graph can be established and a modified deployment can be made integrating the module into the existing system (Figure \ref{fig:fully_deployed}).

\section{CONCLUSION}
This research advances the domain of SDV and centralized vehicle architectures by addressing key challenges within this area. Our contribution is a comprehensive approach towards modular, scalable, and safe centralized automotive systems. By integrating a hardware abstraction layer and a dynamic software deployment system, we have successfully demonstrated that centralized architectures can meet the evolving demands for handling the growing software and hardware complexities in the automotive industry. The modular setup not only streamlines the development and deployment of new applications but also ensures that updates and upgrades can be implemented across different vehicle configurations without extensive modifications.

Our practical evaluation showcased the framework's effectiveness in managing complex automotive operations. The implementation of modules for lane detection, motion planning, and vehicle control further emphasizes the robustness and adaptability of our proposed framework under different software requirements. This work not only contributes to the theoretical aspects of SDV but also demonstrates practical applications that can bridge the gap between current technologies and future automotive software needs.

\section{FUTURE WORK}
In future research on centralized car architectures, several key areas need further exploration. First, enhancing the robustness of the framework through the integration of advanced cybersecurity measures such as encryption, intrusion detection systems, and authentication protocols is crucial, especially since autonomous driving technology advances. This also includes optimizing data throughput and processing speeds for real-time applications, ensuring reliable vehicle operations. Additionally, integrating machine learning approaches could upgrade the framework by enabling systems that dynamically extend themselves. Another important aspect which requires further research is vehicle connectivity. This includes but is not limited to the connection to other traffic participants, the cloud of the vehicle manufacturer, or roadside infrastructure. For future vehicles this communication ensures that software, including underlying frameworks like AUTOFRAME are always up-to-date, so that bug fixes, new features, and safety mechanisms can be automatically deployed to the car without the requirement of complicated interventions e.g. at a car repair shop. By implementing connectivity to external data sources such as roadside infrastructure, the scalability of the framework can be demonstrated at the same time, due to the additional data input apart from the vehicle's sensors. Finally, conducting comprehensive field trials to validate system performance in real-world conditions is essential, to test the system across various environments and operational scenarios.

\printbibliography

\end{document}